\documentclass{article}

\usepackage{PRIMEarxiv}

\usepackage[utf8]{inputenc} % allow utf-8 input
\usepackage[T1]{fontenc}    % use 8-bit T1 fonts
\usepackage{hyperref}       % hyperlinks
\usepackage{url}            % simple URL typesetting
\usepackage{booktabs}       % professional-quality tables
\usepackage{amsfonts}       % blackboard math symbols
\usepackage{nicefrac}       % compact symbols for 1/2, etc.
\usepackage{microtype}      % microtypography
\usepackage{lipsum}
\usepackage{subcaption}
\usepackage{longtable}
\usepackage{supertabular}
\usepackage{fancyhdr}       % header
\usepackage{graphicx}       % graphics
\graphicspath{{media/}}     % organize your images and other figures under media/ folder

\title{A Systematic Literature Review of Game-based Assessment Studies: Trends and Challenges
}

\author{
  Manuel J. Gomez, José A. Ruipérez-Valiente, Félix J. García Clemente \\
  Faculty of Computer Science \\
  University of Murcia \\
  Murcia (Spain)\\
  \texttt{\{manueljesus.gomezm, jruiperez, fgarcia\}@um.es} \\
}

\begin{document}
\maketitle

\begin{abstract}
Technology has become an essential part of our everyday life, and its use in educational environments keeps growing. In addition, games are one of the most popular activities across cultures and ages, and there is ample evidence that supports the benefits of using games for assessment. This field is commonly known as game-based assessment (GBA), which refers to the use of games to assess learners' competencies, skills, or knowledge. This paper analyzes the current status of the GBA field by performing the first systematic literature review on empirical GBA studies. It is based on 65 research papers that used digital GBAs to determine: (1) the context where the study has been applied; (2) the primary purpose; (3) the domain of the game used; (4) game/tool availability; (5) the size of the data sample; (6) the computational methods and algorithms applied; (7) the targeted stakeholders of the study; and (8) what limitations and challenges are reported by authors. Based on the categories established and our analysis, the findings suggest that GBAs are mainly used in K-16 education and for assessment purposes, and that most GBAs focus on assessing STEM content, and cognitive and soft skills. Furthermore, the current limitations indicate that future GBA research would benefit from the use of bigger data samples and more specialized algorithms. Based on our results, we discuss current trends in the field and open challenges (including replication and validation problems), providing recommendations for the future research agenda of the GBA field.
\end{abstract}

% keywords can be removed
\keywords{Game-based assessment \and educational technology \and game-based learning \and learning analytics}

\section{Introduction}
\label{sec:Introduction}

%Contexto
Technology is progressively changing the world in which we live. During the last decade, it has started to make a significant impact on educational environments, and increasing evidence has been accumulated showing the positive impact of technology in education \cite{eiland2019considerations}. One of the most prominent examples of technology in education is the use of digital games \cite{de2018games}. This type of games has become a significant part of families and, especially, among young people around the world. In fact, three-quarters of all U.S. households have at least one person who plays video games \cite{2020gameFacts}, while in Europe, 51\% of the population aged 6-61 years play video games (an average of 8.6 hours/week) \cite{2020gameEuropeFacts}. Moreover, many educators see digital games as powerfully motivating digital environments because of their potential to enhance student engagement and motivation in learning \cite{papadakis2018use}. This increasing interest provides an opportunity to use video games as a tool to improve learning and education. Specifically, there is much enthusiasm in the field of education about game-based assessment (GBA) because conventional assessment methods do not seem to fully have the power to measure all aspects of students’ knowledge, skills, and attributes \cite{de2017future}.

Accompanying this explosion in technology use is the quantity, range and scale of data that can be collected, which have increased exponentially over the last decade \cite{clow2013overview}. In education, the increase in e-learning resources, educational software like Google Classroom or Kahoot, and the use of the Internet have created large repositories that provide a goldmine of educational data that can be explored and used to understand how students learn \cite{romero2013data}. Regarding games, they allow recreating more authentic situations compared to traditional classroom activities, such as lectures or written exercises. From these situations, we can collect a vast amount of detailed data on students' interaction with the game, which provides a great opportunity to make game-based assessments (GBAs) in ways that are not possible in traditional testing \cite{kim2015opportunities}.

% Motivacion: El potencial que tienen los juegos para la evaluación de competencias a través de los datos generados por los estudiantes.
In the past 10 years, numerous studies (see the work in \cite{clark2016digital} for a meta-analysis) have reported that games can be more effective for learning than other traditional teaching methods. In addition, when measuring the competencies acquired, most traditional tests present individual and decontextualized items to learners, while 21st-century competencies benefit from being applied in context for more accurate measurements. Furthermore, classic assessment often interrupts the learning process, and it does little to motivate learners \cite{Dicerbo201317}. Since digital games often employ challenging, interesting, and complex problems, they can be used to generate evidence of 21st-century competencies, which are traditionally difficult to measure using conventional forms of assessment \cite{Ruiperez-Valiente2020648}. The advantages of using games as a form of assessment are manifold \cite{Dicerbo201317, ifenthaler2012assessment, Mislevy201523}: they are engaging and motivating (which provides more valid assessments), and they allow us to create more complex and authentic scenarios required to assess the application of knowledge and skills. Moreover, immediate feedback based on learners' activity can reveal teachers' specific areas of difficulty to make learners keep up with the pace of the class, and such assessment would result in an adaptive game environment, which changes with learners' activity.

% Desafio: ¿Cómo es el estado del arte actual en el area de GBA para su implementación de forma más sistemática?
% Objetivo/contribución: Responder RQs en base a una revisión de la literatura.
The implementation of assessment features into game environments is only in its early stages because it adds a very time-consuming step to the design process \cite{kim2019game}. This situation calls for a review of the current state of the art in the GBA field for effective implementations. In this respect, we found some previous works that performed meta-reviews of the existing research on the different applications of games in learning and education. For example, the authors in \cite{qian2016game} reviewed 137 papers to determine what empirical evidence existed concerning the effects of Game-based Learning (GBL) on 21st-century competencies and identified successful game-design elements that aligned well with established learning theories. Moreover, Alonso-Fernandez et al.\cite{alonso2019applications} carried out a review focused on data science applications to game learning analytics data, showing that the primary purpose when analyzing data from serious games was assessment. Furthermore, Gris and Bengtson \cite{gris2021assessment} aimed to answer how learning, engagement, and  usability  of  games  are  evaluated  in GBL research. To this aim, they conducted a systematic review of 91 empirical studies and categorized their measures and instruments. The researchers concluded that future research in GBL studies should add direct assessments  and  indirect  measures to assess engagement and usability. Guan et al. \cite{guan2022applying} provided a systematic review of 35 experimental studies that substantially integrated gaming elements in primary school lessons and they noted that gamification was the most frequently used game genre. Finally, Chen et al. \cite{chen2021three} conducted a systematic review of 146 articles related to GBL in science and mathematics education. These researchers concluded that GBL is mainly used to increase learner motivation and engagement and reduce learning anxiety. They also revealed that analyzing higher-order thinking skills (e.g., problem-solving, group collaboration) is one of the main hot topics in the community.

Despite the previous reviews of the use of games in learning in education, we have not found any specific study reviewing literature about GBA. For this reason, the current paper aims to conduct the first systematic literature review on the applications of empirical GBA studies and answer some research questions based on the analysis performed to discover current trends and open challenges in this area. The results obtained will provide an overall view of the GBA field, defining its current status and potential future steps in the research in this area. 

% Estructura del paper.
The rest of the paper is organized as follows. Section \ref{sec:methods} describes the methods, including some terminology clarifications, the research questions, databases and search terms, research selection as well as review process. Section \ref{sec:results} presents the analysis and synthesis of our results. Then, we end the paper with a discussion in Section \ref{sec:discussion} and conclusions in Section \ref{sec:conclusion}.

\section{Methods}
\label{sec:methods}

\begin{figure}[ht]
\includegraphics[width=0.7\textwidth]{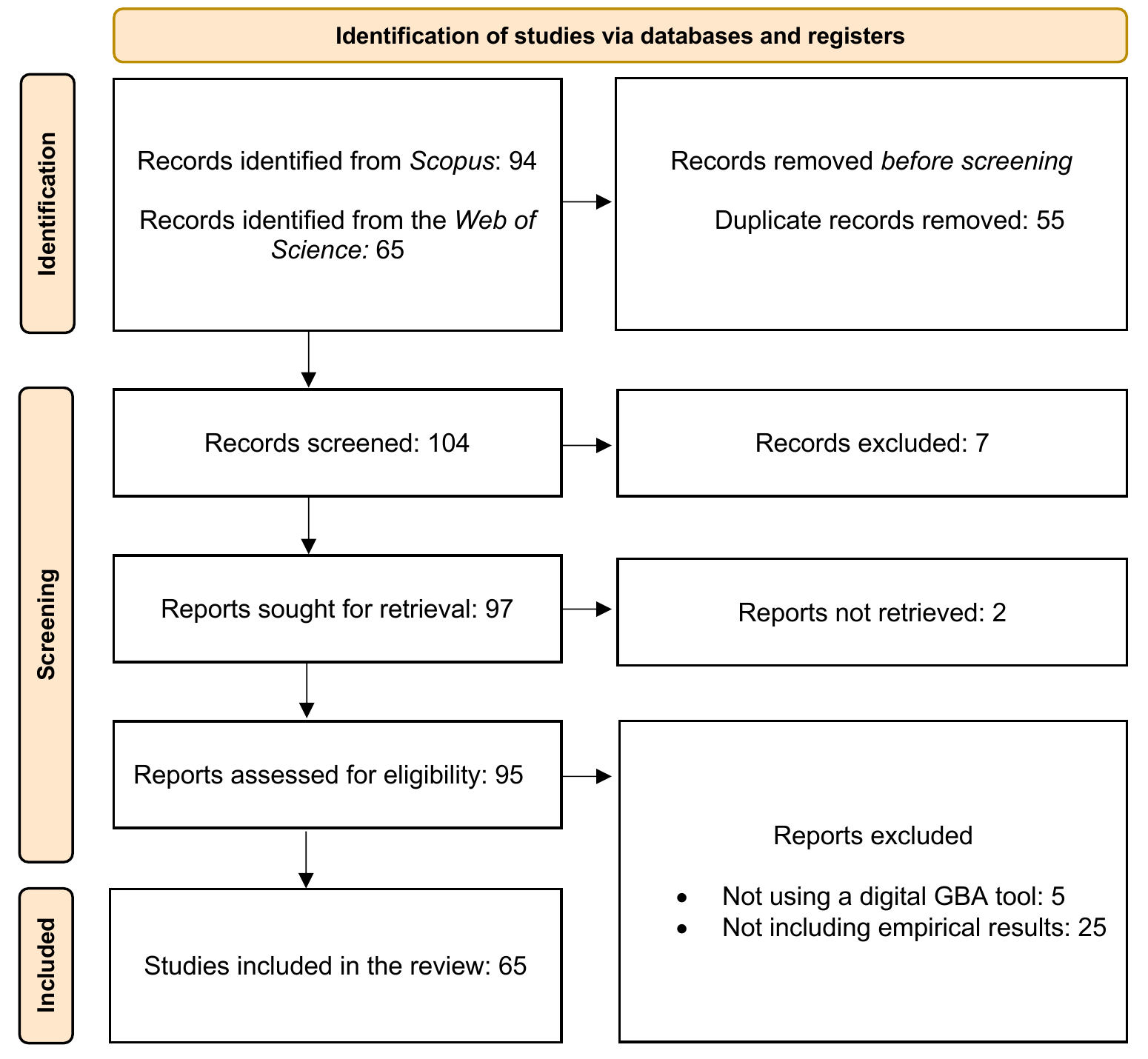} 
\centering
\caption{Flow diagram representing the different phases of the systematic review.}
\label{fig:methodology}
\end{figure}

We followed a standard systematic literature review methodology, using the Preferred Reporting Items for Systematic Reviews and Meta-Analyses (PRISMA) \cite{Pagen160} as a basis for conducting our systematic review. First of all, (1) we formulated each research question (RQ). Then, we used (2) a fixed set of queries on a pre-identified bibliographical database, and (3) a set of inclusion and exclusion criteria. Next, (4) we made a full paper review and coding process of the RQs, and, finally, we carried out (5) a synthesis and analysis. No time restrictions were set. We can see a flow diagram representing the different stages of our systematic review (following the PRISMA template \cite{page2021prisma}) in Figure \ref{fig:methodology}.

\subsection{Terminology clarifications}
\label{subsec:essentialDefs}

In this Section, we present a set of definitions that aim to clarify the concept of GBA, which is the focus of this  systematic literature review. Firstly, we can define a \textbf{game} as ``a system in which players engage in an artificial conflict, defined by rules, that results in a quantifiable outcome'' \cite[p.~80]{tekinbas2003rules}. In addition, games also have clearly-defined goals and obstacles for the player to overcome, providing only intrinsic rewards (satisfaction for getting the right answer) \cite{kang2014interactive}. Secondly, \textbf{GBL} can be regarded as an innovative learning approach where a game is developed to deliver immersive and attractive learning experiences aiming at particular learning goals, experiences and results \cite{de2006learning}. Thus, GBL uses a game containing learning content derived from school curricula or essential life skills to improve the learning experience. Moreover, \textbf{GBA} is a specific application of games, referring to a type of assessment that uses players' interactions with the game, both digital and non-digital, as a source of evidence to make meaningful inferences about what players know and can do (i.e., knowledge, skills), and how individual players interact with the game as a problem-solving process \cite{kim2019game,Kim2016142}. Finally, we have the concept of \textbf{gamification}, which is usually defined as ``the use of game design elements in non-game contexts'' \cite[p.~9]{deterding2011game}. 

Although GBL and GBA are often confused with gamification and gamified assessment, it is undeniable that some differences exist between them. While GBL implies the use of a game developed for learning purposes, gamification utilizes game elements in non-game contexts, not necessarily using full games inside the activities \cite{al2018game}. Thus, GBA also implies the use of a game developed for assessment purposes, using players' interaction with the game as a way to obtain evidence and use this evidence as a form of assessment. Therefore, tools that use gamified activities to assess students' knowledge (e.g., Kahoot, Duolingo) use gamified assessments, and cannot be considered as GBAs. We can also make a clear distinction between GBA and a simple measurement using games since GBA is intended for evaluating players' skills or knowledge based on their interaction with the game. As Ghergulescu and Muntean state, ``measurement represents the process of collecting the information needed for assessment'' \cite[p.~357]{ghergulescu2012measurement}. In other words, measurements are used as evidence to make meaningful inferences about what players know and can do, while measurements using games do not perform that evaluation. These are the definitions that we applied as part of the systematic review screening process to consider a given paper within the GBA field or not, including or discarding that study.

\subsection{Research questions}
\label{subsec:rqs}

\begin{figure}[!ht]
\includegraphics[width=0.9\textwidth]{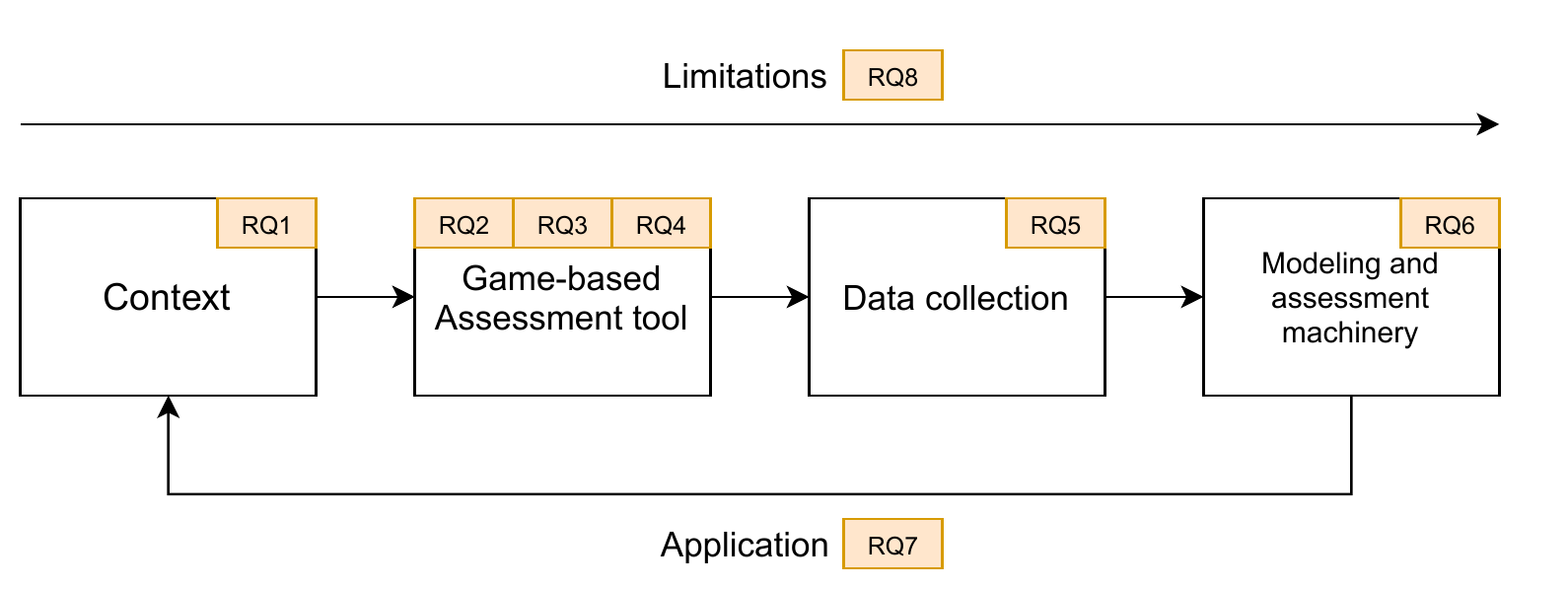} 
\centering
\caption{A simplified view of the steps in game-based assessment empirical research.}
\label{fig:justifRQs}
\end{figure}

To state each one of the RQs, we analyzed and simplified the steps in empirical GBA research \cite{valiente2022unveiling}, which can be seen in Figure \ref{fig:justifRQs}. 

In this process, we can identify five different stages: (1) Learning environments, with the context and learners; (2) The GBA tool that is going to be used in the research; (3) Data collection, to identify which data has to be collected and how to store them; (4) Modeling and assessment machinery; and (5) Educational application, to identify the final objective and target users. From these stages, we identified the following RQs, which allow us to understand the open challenges and current trends in the area:

\begin{enumerate}
    \item[\textbf{RQ1.}] In what context or environment has GBA been applied?
    \item[\textbf{RQ2.}] What is the primary purpose of GBA?
    \item[\textbf{RQ3.}] What is the domain of GBA?
    \item[\textbf{RQ4.}] Is the game/tool used available to the public?
    \item[\textbf{RQ5.}] What is the size of the data sample used in the study?
    \item[\textbf{RQ6.}] What computational methods and algorithms have been applied in the research?
    \item[\textbf{RQ7.}] What stakeholder is the intended recipient of the research results?
    \item[\textbf{RQ8.}] What limitations and challenges do the authors address?
\end{enumerate}

In addition, Figure \ref{fig:justifRQs} also shows the mapping between the different stages and the RQs identified: RQ1 is based on the first stage, related to the learning context. Then, RQ2, RQ3, and RQ4 are based on the second stage, which refers to the GBA tool used, its primary purpose, the domain and availability. Next, we wanted to investigate the sample size (RQ5) which is situated in the data collection stage. Regarding the modeling and assessment machinery, our objective was to investigate the computational methods and algorithms applied in the research (RQ6). RQ7 refers to the application of the research in the desired context, identifying the main stakeholder of the research. Finally, RQ8 aims to identify the research limitations at any stage.

\subsection{Databases and search terms}
We have queried two databases: Scopus and the Web of Science since they are the most widely used databases in different scientific fields and are often used for surveying the literature \cite{aghaei2013comparison}. Scopus is the world's largest citation database of peer-reviewed research literature, with over 22,000 titles (including journals, conferences and book series) from more than 5,000 international publishers, of which 20,000 are peer-reviewed journals in the scientific, technical, medical, and social sciences \cite{scopus}. Moreover, the Web of Science, the second biggest bibliographic database, can be used to track ideas going back several decades from almost 1.9 billion cited references from over 171 million records \cite{wos}.

To perform the search on both databases, we restricted the query to title and keywords: 1) we included the term ``game-based assessment'' and searched for it within the paper titles; 2) we included the term ``game-based assessment'' and searched for it within the paper keywords. Thus, we used the following final search query:

\textit{(TITLE}(``game-based assessment'') OR \textit{KEY}(``game-based assessment'')).

The initial selection of studies was retrieved in January 2021, and this query generated 159 initial studies (94 from Scopus and 65 from the Web of Science).

\subsection{Inclusion/Exclusion criteria}

After obtaining the initial collection, we excluded the duplicated records from the two databases (55 studies). Then, we made a first brief review of all papers, comparing them against the inclusion and exclusion criteria. This first review was conducted by one of the authors. After the first analyses, we classified studies as \textit{included} or \textit{excluded}, and the coding results were discussed collaboratively by the three authors in order to obtain the final set of included and excluded studies and avoid possible errors. The inclusion/exclusion criteria followed are described below. Given these criteria, the paper was included if all of the following conditions were met (i.e., if one condition was not met, the paper was excluded). Furthermore, the conditions were applied sequentially, so that a paper not matching a condition was excluded immediately from the collection:

\begin{itemize}
    \item The paper was written in English or Spanish (languages in which the authors have high proficiency): 0\% of the papers were excluded.
    \item The paper was fully accessible: 1.9\% of the papers were excluded (2 studies).
    \item The paper was published in conference proceedings, journals or edited books/volumes (i.e., book chapter): 0\% of the papers were excluded.
    \item The paper was not extended at a later time (i.e., a conference paper that was later on extended in a journal paper): 6.9\% of the papers were excluded (7 studies).
    \item The paper used a digital GBA tool: 5.3\% of the papers were excluded (5 studies). See Section \ref{subsec:essentialDefs} above for relevant definitions.
    \item The paper included empirical evidence related to the outcomes of applying the GBA tool: 27.8\% of the papers were excluded (25 studies).
\end{itemize}

\subsection{Final paper collection}

After the first brief review to ensure that every paper met our inclusion/exclusion criteria, we excluded a total of 39 papers. Thus, the final paper collection consists of 65 studies.

\begin{figure}[ht]
\includegraphics[width=0.9\textwidth]{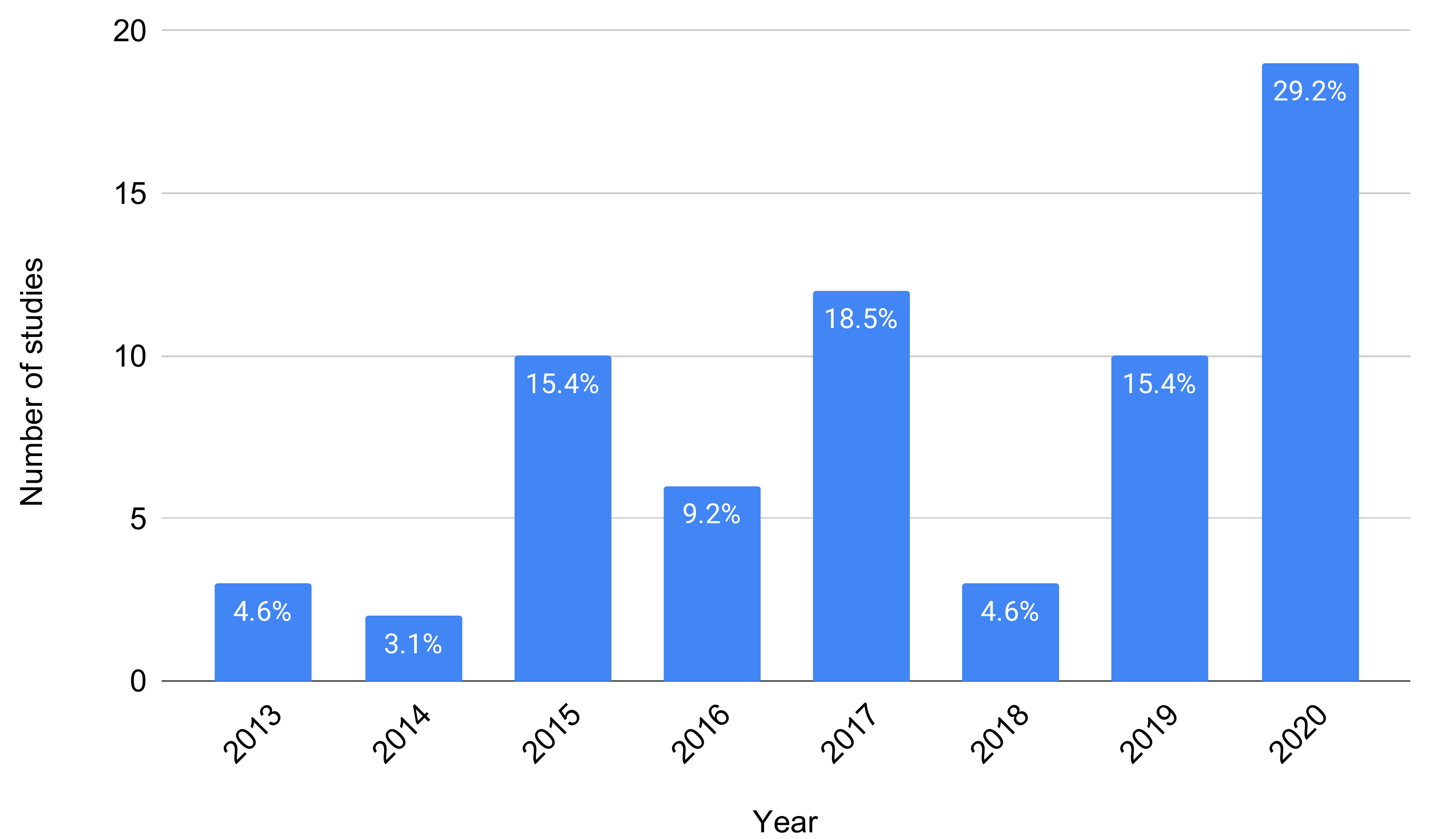} 
\centering
\caption{Number of selected studies per year of publication.}
\label{fig:paperYear}
\end{figure}

Figure \ref{fig:paperYear} shows the distribution of papers within the final collection by publication year. We see an increasing interest in this particular topic: between the years 2013 and 2016, we only have 21 (32.3\%) published papers that matched our criteria, while between 2017 and 2020 there are 44 (67.7\%) of them.

\begin{figure}[ht]
\includegraphics[width=0.9\textwidth]{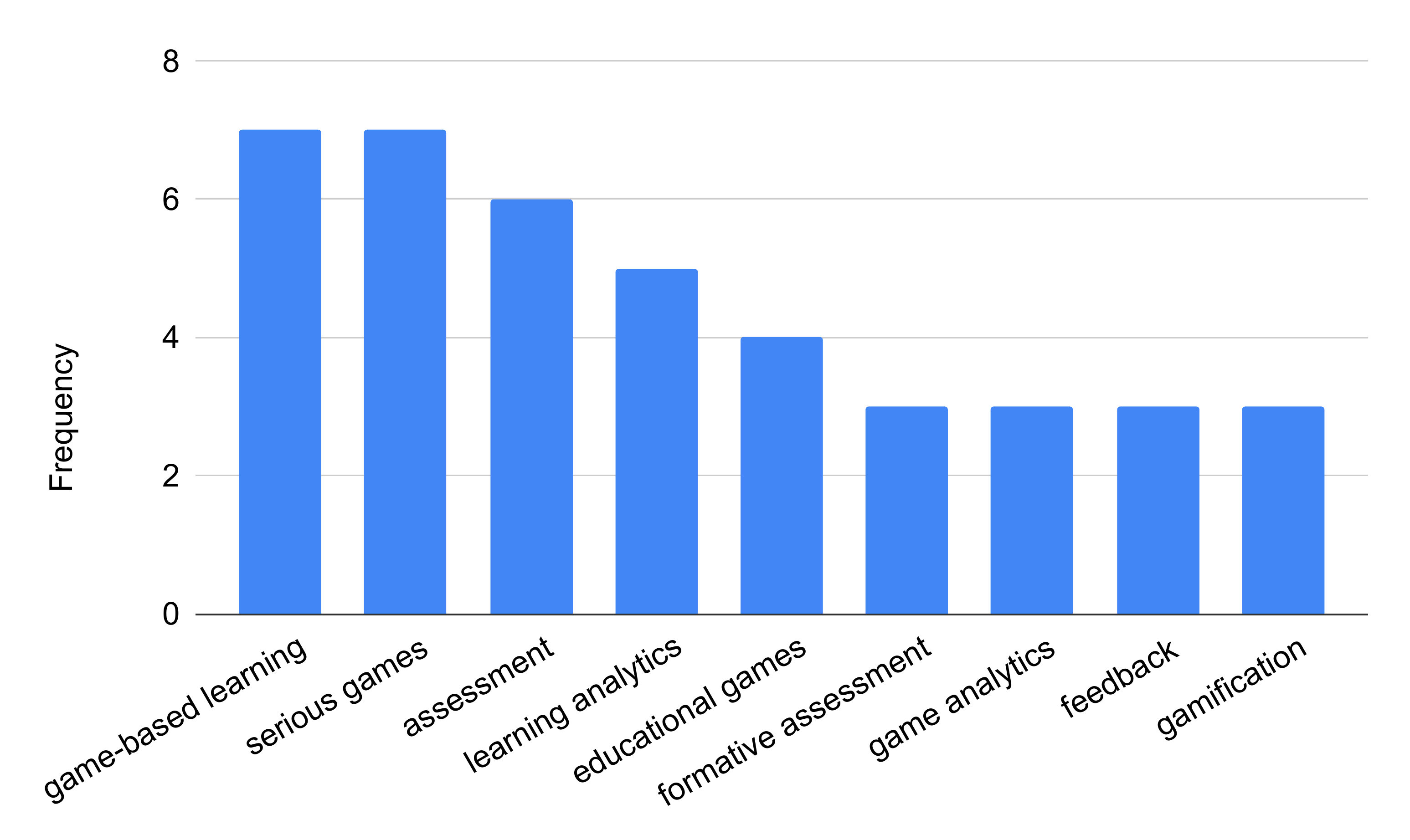} 
\centering
\caption{Distribution of keywords across articles in the final collection.}
\label{fig:paperKey}
\end{figure}

We also collected each paper's keywords and made a brief analysis to describe our paper collection. For our analysis, we excluded the ``game-based assessment'' keyword since it was the most common one. The most frequent keywords are presented in Figure \ref{fig:paperKey}. The total sum of keywords is 299 while there are 201 unique keywords. The average keyword was found 1.48 times. As we can see, the predominant keywords strongly focused on games, assessment, and analytics.

\subsection{Review and coding process}

In the coding stage, we collected the data of the selected studies that we consider to be the most valuable to address the RQs in Section \ref{subsec:rqs}. Based on the aim of the review, we followed an inductive coding scheme (also called open coding). This means that the codes created were based on the qualitative data itself \cite{inductiveCoding}. This is an iterative process since researchers can add new codes, split an existing code into two, or compress two existing codes into one as they continue reviewing data. Specifically, in our analysis, we first made a brief review of each paper (conducted by one author), collecting all the necessary information to code each RQ at once. After that, we followed an iterative process whereby we continued reviewing the information corresponding to each RQ sequentially, and unclear results were discussed and contrasted by the three authors. The full results of the coding process per paper are available in \cite{paperCoding}. In addition, it should be noted that each paper can fit into more than one of the codes created for each RQ.

\section{Results}
\label{sec:results}

\subsection{In what context or environment has the GBA been applied? (RQ1)}
\label{subsec:rq1}

GBAs can be used in very different environments. Our analysis reveals that there are three main contexts where GBAs have been used:

\begin{enumerate}
    \item \texttt{K-16 education}: some papers use GBAs in K-16 education (e.g., school, university) to support teaching and learning. More specifically, games are most commonly used in middle school and high school (23.1\%). However, games are also used in other K-16 education environments such as primary school (15.4\%) and university (10.8\%). For example, Di Cerbo et al.\cite{DiCerbo2015319} used game data from 751 US middle school players.
    
    \item \texttt{Medical}: games can also be used in medical environments for different purposes (e.g., rehabilitation). For example, the authors in \cite{Lindenmayer2020166} examined the feasibility of administering the GBA in a sample of inpatients with chronic schizophrenia with low levels of functioning. Moreover, the authors in \cite{wiloth2016validation} aimed to present data on construct validity, test–retest reliability and feasibility, measuring motor-cognitive functions in multimorbid patients with mild-to-moderate dementia. Regarding construct validity, the authors tested eight hypotheses and confirmed seven of them (87.5\%), thus indicating excellent construct validity. Moreover, the authors found moderate-to-high test–retest reliability (ICC=0.47-0.92).
    
    \item \texttt{Workforce}: another option is the use of games to assess in professional environments. In this context, enterprises can use games to evaluate their employees or provide them with additional feedback. Even now, companies can include the use of GBA for the recruitment of staff and the selection process \cite{collmus2016game}. This idea is supported by the fact that in-game constructs show similar relationships with in-game performance to what the workforce constructs do with job performance \cite{Short2019161}.
\end{enumerate}

% Grafico con arbol en vez de las dos graficas
%figure 
\begin{figure}[!ht]
    \includegraphics[width=0.9\textwidth]{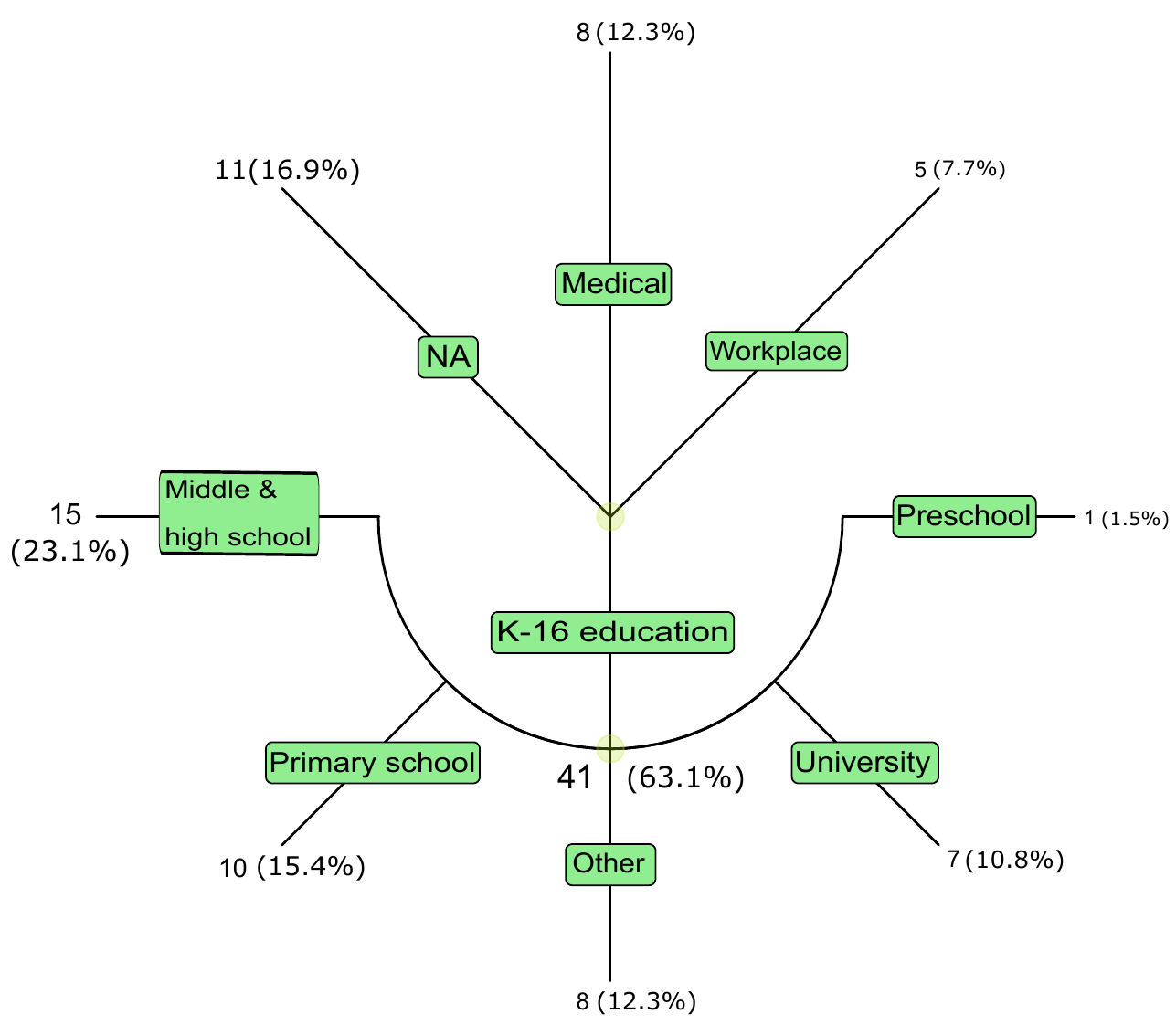} 
    \centering
    \caption{Papers' category distribution based on RQ1.}
    \label{fig:rq3fig}
\end{figure}

There are also some studies such as \cite{stanciu2019development} that did not specify in what context their games were used (11 studies, 16.9\%). We can see the number of papers fitting in each category in Figure \ref{fig:rq3fig}. As the figure shows, GBAs are mostly used in \texttt{K-16 education} (41 studies, 63.1\%), followed by \texttt{medical} (8 studies, 12.3\%) and \texttt{workforce} contexts (5 studies, 7.7\%).

\subsection{What is the primary purpose of the GBA research? (RQ2)}
\label{subsec:rq2}

In this RQ, we wanted to know what was the main purpose of each GBA in each study. We coded the papers' main purpose into six different categories: \texttt{GBA evaluation}, \texttt{study of in-game behaviors}, \texttt{assessment}, \texttt{interventions}, \texttt{framework proposal} and \texttt{game design proposal}. Next, we describe in detail each one of these categories.

\begin{enumerate}
    \item \texttt{GBA evaluation}: in these studies, authors evaluate the game by checking if it achieves its initial objectives using some measure to prove that the game or tool is suitable for an educational environment. In \cite{Hummel2017225}, the authors showed how they applied the methodology for an assessment game for ICT managers in secondary vocational education, checking if this assessment was content-valid compared to a face-to-face assessment. Moreover, the authors in \cite{marengo2020innovative} aimed to investigate whether it is possible to perform an in-Basket test (which is widely used by companies and organizations in order to map employees' soft skills) online with the same effect as that of the onsite one.
    
    \item \texttt{In-game behaviors}: in these studies, authors investigate in-game players' behaviors (e.g., persistence, engagement). By identifying these behaviors, we can group players according to different behaviors or simply check if a student shows a specific one. For example, Dicerbo \cite{Dicerbo201317} used evidence extracted from log files to create a measure of persistence. Similarly, Ventura \& Shute \cite{Ventura20132568} also created a measure of persistence, validating it against another existing measure and concluding that the GBA predicted students' learning.
    
    \item \texttt{Assessment}: in these studies, games are used to report measures that aim to evaluate students. This allows for improvements in the learning process using this evaluation measure instead of classic evaluation methods or providing personalized feedback. In their work, Weiner \& Sanchez \cite{Weiner2020215} created an alternative measure using a virtual reality game that calculated scores to indicate specific cognitive abilities.
    
    \item \texttt{Interventions}: games can also be used to investigate the effect of some interventions while playing. For example, we can use feedback messages to notify the learner with positive (or negative) feedback to observe how this intervention influences its performance and behavior. Another typical example is switching the order of in-game elements or testing different game features. In \cite{Cutumisu2019}, the authors used a psychophysiological methodology to investigate attention allocation to different feedback valences (i.e., positive and negative feedback). With that purpose in mind, they used an eye tracker to collect accurate information about individuals' locus of attention when they process feedback. 
    
    \item \texttt{Framework proposal}: in these papers, the authors propose the design of a new framework to be used within the context of GBA. We can see an example in \cite{Almond2015250}, where the authors examined the process of creating a Bayesian network framework through different techniques (e.g., using correlation matrixes, IRT) to create scoring models for assessing students.
    
    \item \texttt{Game design proposal}: authors provide a game design that can be used for assessment purposes. For example, the authors in \cite{Rivera2015} show the design of an online GBA to help students improve their learning outcomes and promote the development of general and transferable skills, such as the ability to solve problems in complex situations, and working under pressure. 
\end{enumerate}

Some studies focused on more than one of the categories described above. For example, Weiner \& Sanchez \cite{Weiner2020215} used a virtual reality game to calculate a score measure for each student (\texttt{assessment}) and they proved that these calculated scores are best used by comparing them to classic measures (\texttt{GBA evaluation}). 
 
\begin{figure}[ht]
\includegraphics[width=0.9\textwidth]{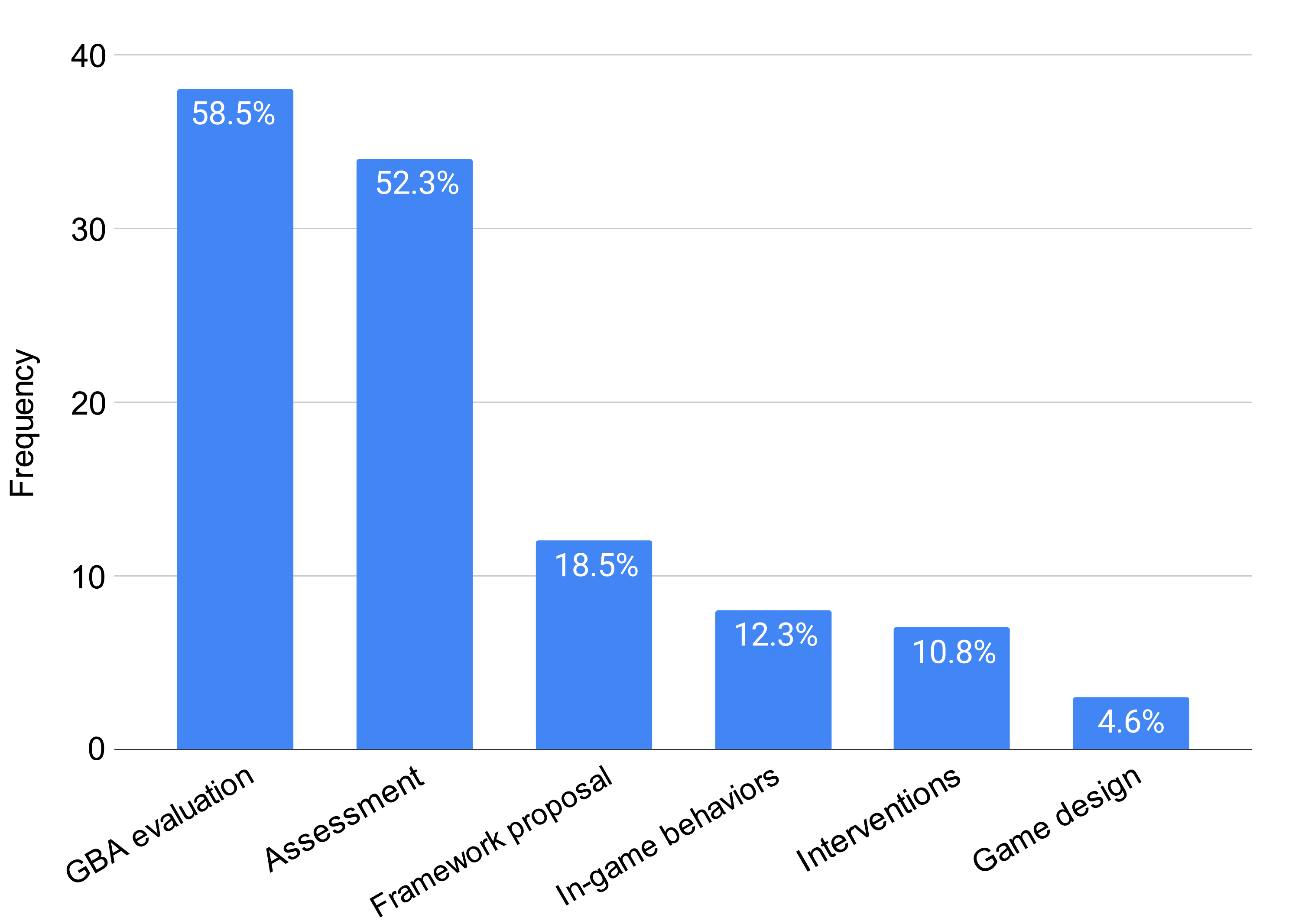} 
\centering
\caption{Main purpose of each research. More than one purpose is possible for each paper.}
\label{fig:purposes}
\end{figure}

We can see the number of papers fitting each category in Figure \ref{fig:purposes}. \texttt{GBA evaluation} is the most common category (38 studies, 58.5\%), followed by \texttt{assessment} (34 studies, 52.3\%) and \texttt{framework proposal} (12 studies, 18.5\%). The less common category is \texttt{game design proposal}, with only three papers fitting (4.6\%). We can conclude that most papers focused on using games to assess learning, but they also tried to prove that this assessment was a valid measure to be used in real educational contexts.

\subsection{What is the domain of the GBA? (RQ3)}
\label{subsec:rq3}

From reviewing the selected papers, we identify four major domain categories: \texttt{STEM}, \texttt{humanities and social sciences}, \texttt{cognitive and soft skills}, and \texttt{physiological capacities}. As some of the categories have more than one related area, we also consider some sub-categories fitting into them. We describe each domain category below in detail:

\begin{enumerate}
    \item \texttt{STEM}: in this category, we include papers that are related to \texttt{science}, \texttt{technology}, \texttt{engineering} and \texttt{mathematics}. For example, Chiu \& Hsieh \cite{Chiu20171075} showed the different teaching methods of second-grade elementary students in fraction concepts (\texttt{mathematics}), while Kim et al. \cite{Kim2016142} aimed to assess the understanding of Newton’s three laws of physics using a two-dimensional physics game.
    
    \item \texttt{Humanities and social sciences}: papers related to humanities and social science areas (e.g., art, music, language) fit in this category. As this is a wide area, we have also defined some sub-categories to better categorize the papers. These sub-categories are \texttt{language}, \texttt{art} and \texttt{history}. Studies that do not fit into one of those three categories are categorized as \texttt{other}. As an example of the \texttt{art} category, we highlight the work in \cite{Basu2020985}, where the authors used a game in which players collect data about the musical interests of an in-game character and use these data to make decisions about which artists to sign and what songs to record. We can see another example (related to \texttt{language}) in \cite{Song20191324}, where the researchers described the design of an argumentative reasoning task within a scenario-based assessment enhanced with game elements.
    
    \item \texttt{Cognitive and soft skills}: cognitive skills are the core skills your brain uses to think, read, learn, remember, reason, and pay attention \cite{cognitiveskills}. Cognitive skills help to process new information by taking that information and distributing it into the appropriate areas in the brain. Developing cognitive skills helps to complete this process more quickly and efficiently, helping people to understand and effectively process new information \cite{cognitiveskillsExp}. Moreover, soft skills are described as a combination of interpersonal and social skills, including the ability to communicate, coordinate, work under pressure, and solve problems \cite{dixon2010importance}. In this category, we consider \texttt{attention}, \texttt{memory}, \texttt{logic and reasoning}, \texttt{visual processing and speed}, and \texttt{soft skills}. We find papers that have measured interesting skills, such as \cite{Song20191175}, which included a series of reasoning activities to measure argumentation skills (which is related to \texttt{logic and reasoning}), or \cite{Nikolaou2019, mosalam2019assessing}, which used GBAs to assess candidates’ soft skills.
    
    \item \texttt{Physiological capacities}: physiological functional capacity is the ability to perform the physical tasks of daily life and the ease with which these tasks can be performed. We could assess daily physical tasks, like Rodríguez de Pablo et al. \cite{Rodriguez-de-Pablo2017413}, who used a set of games to provide a fast, quantitative and automatic evaluation of the arm movement function. Furthermore, other works focused on assessing mental abilities, such as motivating children with autism to make more eye contact \cite{Korhonen2017281}.
    
\end{enumerate}

\begin{figure}[ht]
\includegraphics[width=0.9\textwidth]{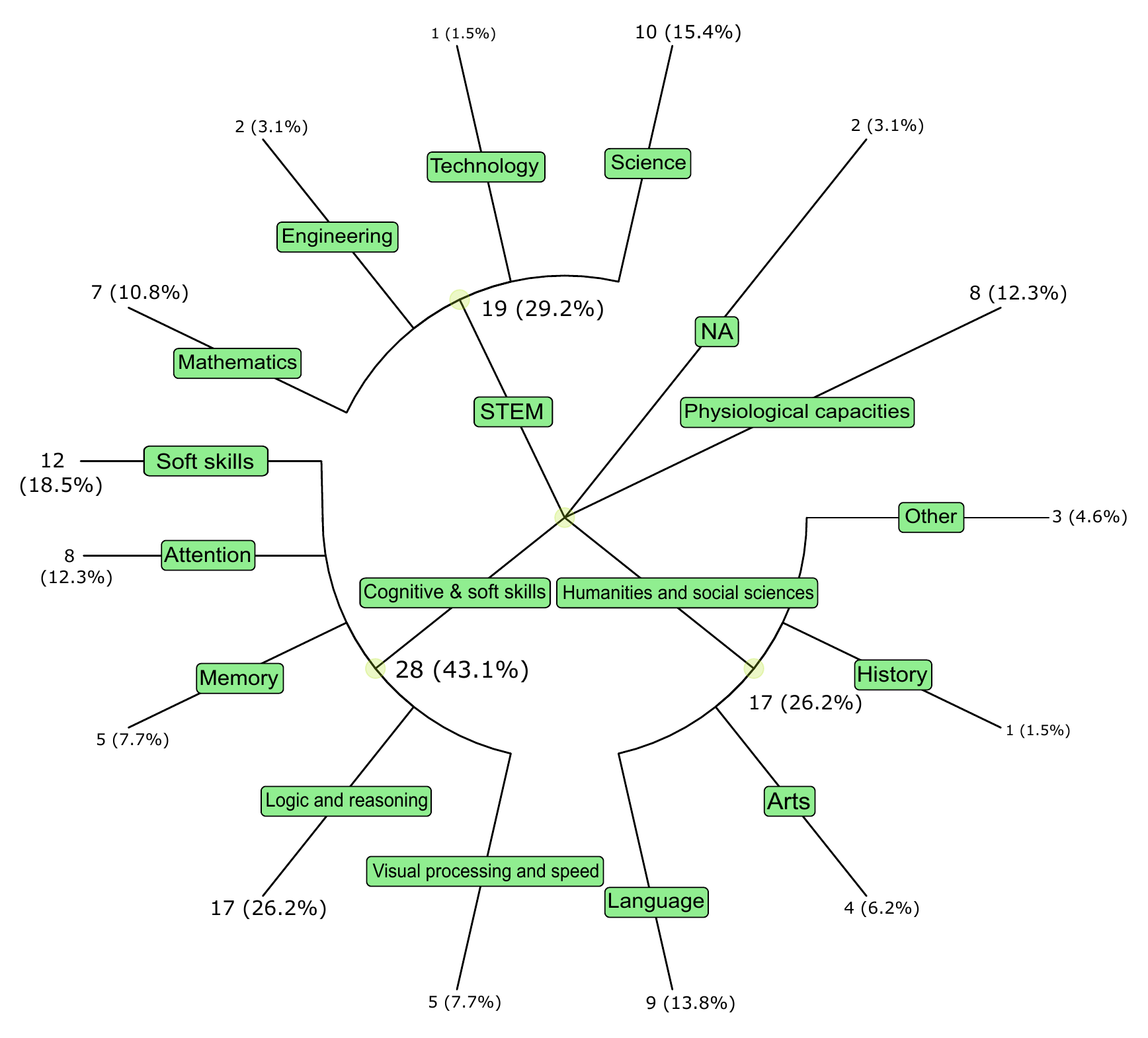}  
\centering
\caption{Category tree for RQ3.}
\label{subfig:rq2tree}
\end{figure}

There are also papers fitting more than one category at once. For example, in \cite{TannerJackson2020, Jackson2018795}, researchers used a GBA for measuring argumentation and pragmatic skills. This research measured language competencies (which is part of \texttt{humanities and social sciences}), but it also measured \texttt{cognitive and soft skills}. We can see the full tree showing the distribution of studies into categories and sub-categories in Figure \ref{subfig:rq2tree}.

The three predominant categories are \texttt{cognitive and soft skills} (28 studies, 43.1\%), \texttt{STEM} (19 studies, 29.2\%), and \texttt{humanities and social sciences} (17 studies, 26.2\%). Taking a look at each sub-category, we note that the main area in \texttt{STEM} is \texttt{science} (10 studies, 15.4\%). The main field in \texttt{cognitive and soft skills} is \texttt{logic and reasoning}, with 17 papers (26.2\%), while in \texttt{humanities and social sciences}, the predominant sub-category is \texttt{language}, with nine papers (13.8\%).

\subsection{Is the game/tool used available to the public? (RQ4)}
\label{subsec:rq4}

A critical aspect of research is the availability of the results obtained to be used by the general public. It is essential to make tools accessible so that researchers can replicate experiments and practitioners can use them as part of their teaching. From our analysis, we find three primary categories: \texttt{Currently available}, \texttt{Not available (NA)} and \texttt{Not specified}. 

\begin{enumerate}
    \item \texttt{Currently available}: the game/tool used in the corresponding research is currently available (using the web portal specified by the authors) for public use (e.g., \cite{Song2020, Ketamo2014258}).
    \item \texttt{Not available}: the game/tool used in the research was presented as initially available in the paper, but currently, it is no longer accessible based on our attempt to access the site (e.g., \cite{Halverson2014111, Gaggi2017155, Bertling2015545}).
    \item \texttt{Not specified}: researchers did not specify the tool's availability; it is more than likely that it is not accessible (e.g., \cite{Baron20151037, Loachamin-Valencia2017106, pouezevara2019assessing}).
\end{enumerate}

\begin{figure}[!ht]
\includegraphics[width=0.9\textwidth]{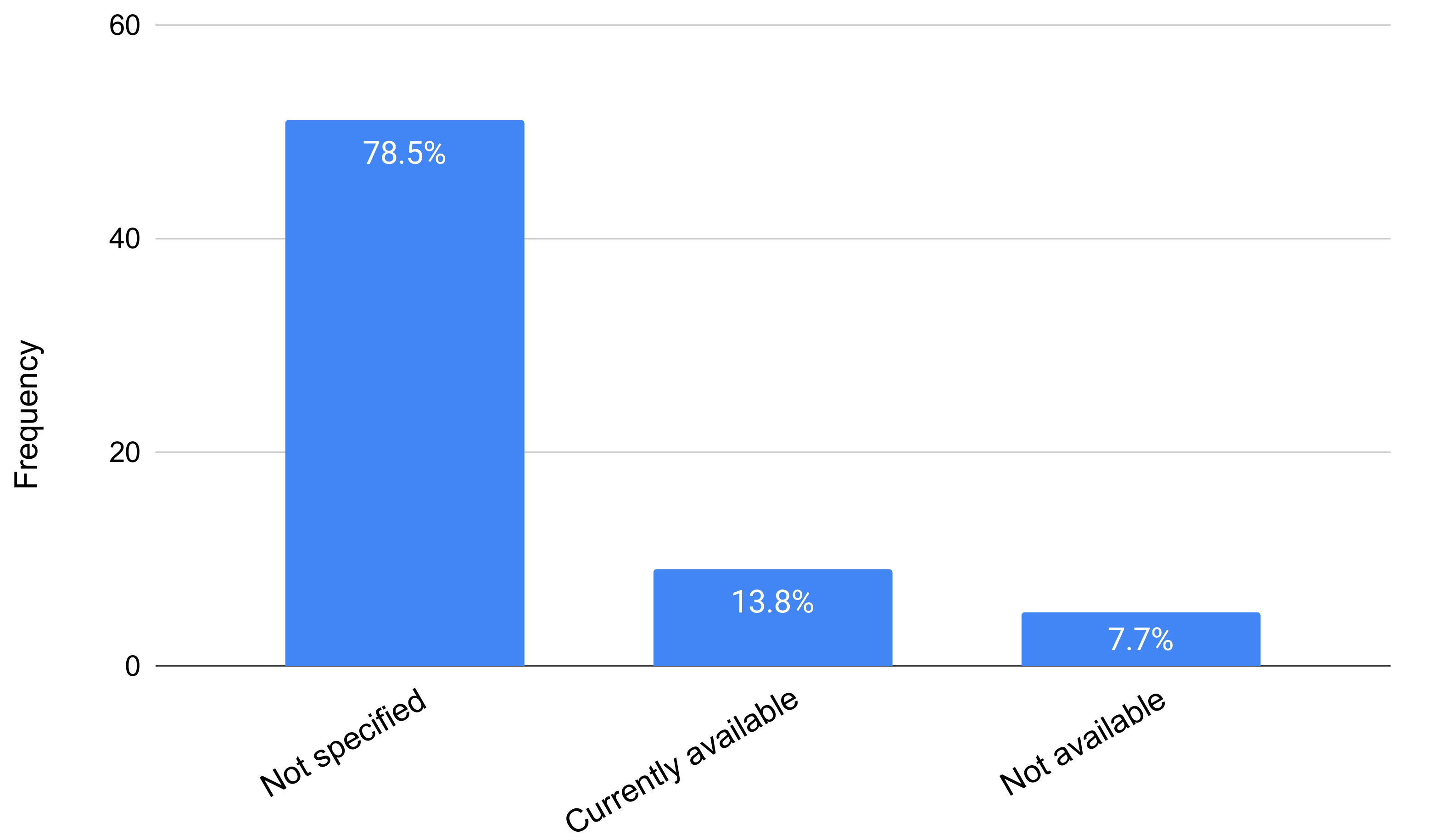} 
\centering
\caption{Number of papers based on their availability.}
\label{fig:rq7}
\end{figure}

Figure \ref{fig:rq7} shows that most papers (51 studies, 78.5\%) did not specify if the tool is accessible or not. Another minority of papers (9 studies, 13.8\%) offered their tool publicly. The rest of the studies (five papers, 7.7\%) initially offered their games, but they are currently unavailable. In addition, we did not find any open-source game across the studies included in the review.

\subsection{What is the size of the data sample used in the study? (RQ5)}
\label{subsec:rq5}

In this research question, we classify the different data collections used in the studies based on their data sample size. Investigating the sample size is relevant since the use of larger data samples will allow better generalization of the research results, as well as the possibility of applying more complex algorithms (e.g., neural networks), which often require large amounts of data to outperform other models \cite{alonso2019applications}. Although the sample size can be relevant for some aspects, such as preventing overfitting in some methods, it is not related to the study's rigor (i.e., using a larger sample does not make a study more rigorous). From the coding process, we present four categories:

\begin{enumerate}
    \item \texttt{Fewer than 50 participants}: these papers involved fewer than 50 participants in their empirical studies. We find studies with small data samples, such as \cite{Mavridis2017137}, using data from 30 postgraduate students, or \cite{Vallejo2017}, which used a sample of 20 healthy controls patients and 18 patients with Alzheimer's disease to evaluate the usability of a tool created to assess cognitive functions.
    
    \item \texttt{Between 50 and 250 participants}: these papers involved between 50 and 250 participants in their studies. For example, Leonardou et al. \cite{Leonardou2020} used data from 77 primary school pupils for assessing and improving multiplication skills. We see another example in \cite{Abeele2015331}, which used data from 95 children from the final year in preschool to measure psychoacoustic thresholds.
    
    \item \texttt{Between 250 and 500 participants}: these papers involved between 250 and 500 participants in their studies. For example, Gjicali et al. \cite{Gjicali2020} used data from 433 students who played a game simulating an artificial culture with norms embodying two cultural concepts: hierarchy and collectivism.
    
    \item \texttt{More than 500 participants}: these papers used data from more than 500 participants in their research. Hautala et al. \cite{Hautala20201003} used data from 723 students to investigate reading difficulties, concluding that the GBA could be successfully used to identify students with reading difficulties with acceptable reliability (Cronbach’s alpha 0.93 and 0.87). Some other studies used a huge sample, such as \cite{Ruiperez-Valiente2020648}, which used data from 5,545 students to measure engagement and cluster students to finally report four different engagement profiles.

\end{enumerate}

\begin{figure}[ht]
\includegraphics[width=0.9\textwidth]{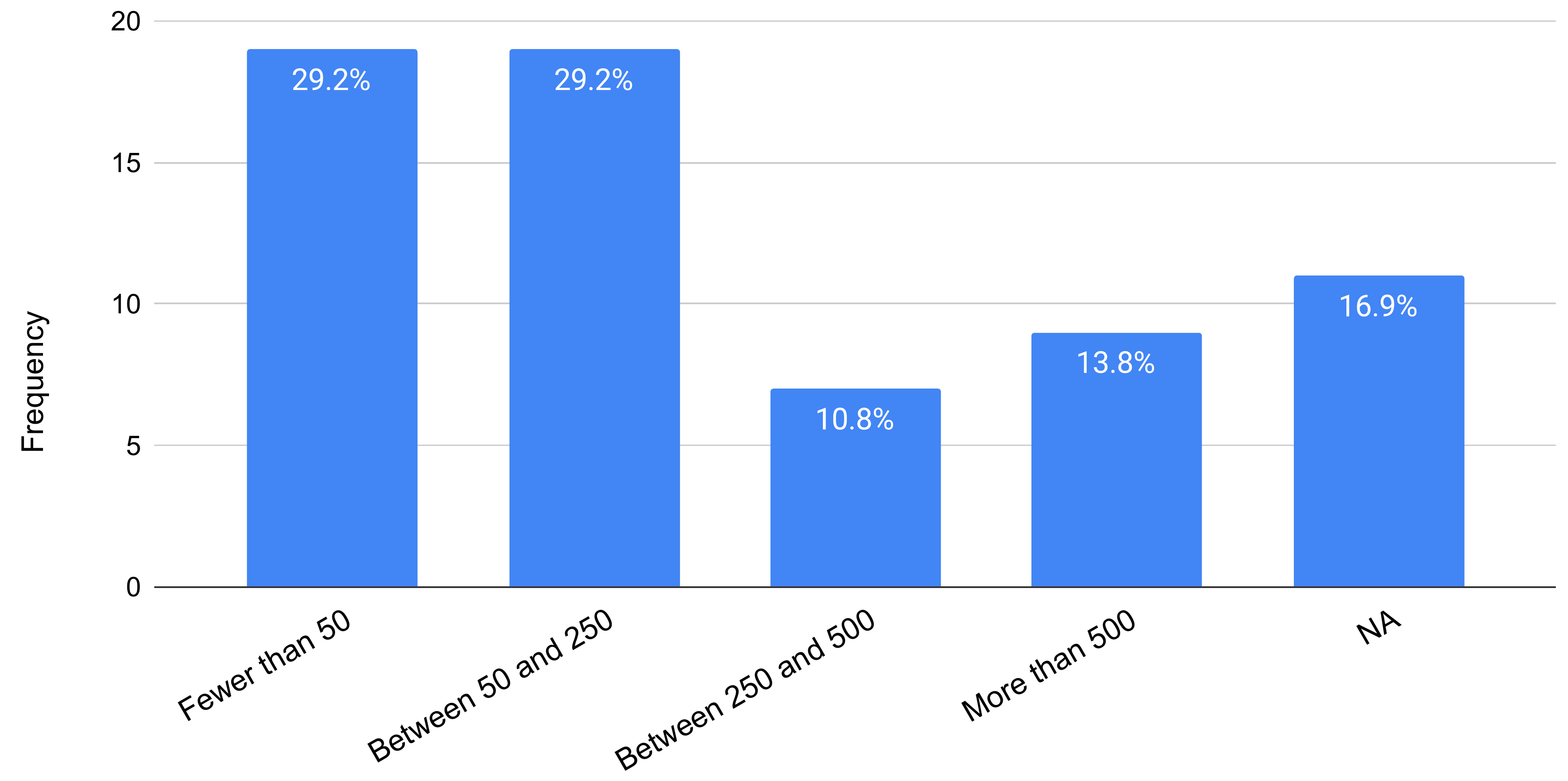} 
\centering
\caption{Paper distribution based on the number of participants involved in empirical studies.}
\label{fig:rq4}
\end{figure}

Other papers (e.g., \cite{Kim2015340, Perry2013}) did not specify the data sample size of the study and we categorize them as \texttt{NA}. Figure \ref{fig:rq4} summarizes the results of the data sizes across papers. We can see that only 16 papers (24.6\%) used more than 250 participants in their students, and only nine papers (13.8\%) used data from more than 500 students, meanwhile most of the papers (58.5\%) used data from fewer than 250 participants. We also see that a significant amount of papers (11 studies, 16.9\%) did not specify the data sample size in their studies.

\subsection{What computational methods and algorithms have been applied in the research? (RQ6)}
\label{subsec:rq6}

After exploring the data samples that were retrieved across papers, our goal was to examine the methods that were applied for its analysis. We believe that the methods being applied are crucial, since they are the link between the evidence generated by learners and the assessment. Accordingly, we identified five different groups of methods for analyzing the data: \texttt{Descriptive statistics}, \texttt{Machine learning}, \texttt{Knowledge inference}, \texttt{Deep learning}, and \texttt{Sequence mining}. Below is a description of each group in detail:

\begin{enumerate}
    \item \texttt{Descriptive statistics}: these encompass further mathematical analyses covering various methods, tests, and visualizations. We identified several papers that applied \texttt{summary statistics} (e.g., mean, variances) \cite{Gomez-Alvarez2017633}, \texttt{correlations} \cite{Jaffal2015101} and \texttt{visualizations} \cite{Cutumisu20192977}.  
    \item \texttt{Machine learning}: it is a part of AI and covers a set of methods that allow systems to learn and improve from historical data automatically. We noted that the authors used two significant families of machine learning methods: \texttt{supervised learning} and \texttt{unsupervised learning}. \texttt{Supervised learning} includes techniques such as regression \cite{Chin2016195} while \texttt{unsupervised learning} uses other methods, such as clustering techniques like \textit{k}-means \cite{Ruiperez-Valiente2020648} or dimensionality reduction techniques like Principal Component Analysis (PCA) \cite{Forsyth2017502}.
    
    For example, the authors in \cite{Arce-Lopera2020243} developed a game for evaluating the logic abilities of first-year university students. They tried to compare the measures obtained by paper-based tests with those obtained using the game by conducting a linear regression (which is a supervised method). The authors concluded that the measures obtained from both methods were not significantly different.
    
    \item \texttt{Knowledge inference}: it refers to the acquisition of new knowledge from existing facts based on certain rules and constraints. One way of representing these rules and constraints is through the use of logic rules, formally known as knowledge representation \cite{Tari2013}. Common knowledge inference methods that several studies have used are \texttt{Bayesian networks} \cite{Shute2016535} and \texttt{fuzzy cognitive maps} \cite{Baron20151037}.
    
    In \cite{Levy2019771}, the researchers proposed a dynamic Bayesian network modeling approach for measuring student performance from an educational video game. The results supported the usefulness of Bayesian networks to characterize and accumulate evidence regarding students in games and related assessment environments.
    
    \item \texttt{Deep learning}: an artificial intelligence function that imitates the workings of the human brain in processing data and creating patterns for decision-making \cite{deeplear}. An example is the work of Chen et al. \cite{Chen2020481}, who used Long Short-Term Memory (LSTM), an artificial recurrent neural network architecture.
    
    \item \texttt{Sequence mining}: the objective of sequence mining is to unlock useful knowledge hidden in sequence data \cite{dong2007sequence}. Specifically, Gomez et al. used \cite{gomez2020exploring} sequence mining to identify sequences and errors by transforming raw data into meaningful sequences that are interpretable and actionable for teachers.
    
\end{enumerate}

\begin{figure}[ht]
\includegraphics[width=0.9\textwidth]{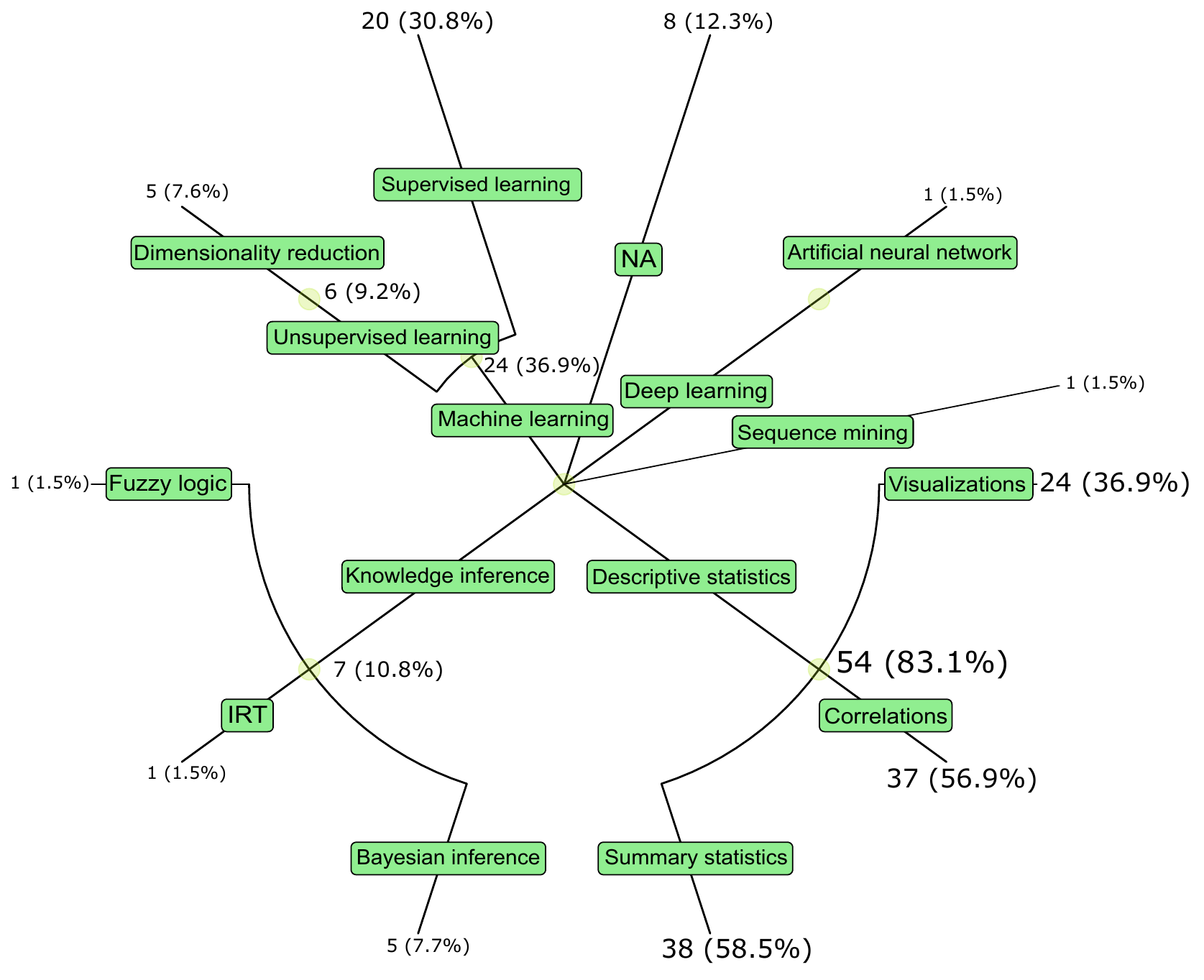} 
\centering
\caption{Paper distribution based on methods used.}
\label{fig:rq5}
\end{figure}

In Figure \ref{fig:rq5} we can see the different families of techniques and the number of papers that used them in their research. We can see that most papers (83.1\%) used \texttt{descriptive statistics}, and almost none of them used \texttt{deep learning} (only one paper, 1.5\%). We also noted that 83.3\% of the papers that used \texttt{machine learning} techniques used \texttt{supervised learning} too, specifically, most of them used different types of regressions.

\subsection{What stakeholder is the intended recipient of the research results? (RQ7)}
\label{subsec:rq7}

A stakeholder is defined as a person with an interest or concern in something, especially a business \cite{oxfordDict}. In our study, we consider the paper's stakeholder as the person to whom the results are directed, even though the paper's contribution might have other secondary stakeholders. Specifically, we have two main groups of stakeholders: \texttt{researchers} and the \texttt{final user}. 

\begin{enumerate}
    \item \texttt{Researchers}: if the paper's contribution is methodological, we expect that the paper's main stakeholders will be researchers. For example, Lonergan et al. \cite{Lonergan2020250} created a Paper-based Assessment (PBA) and a GBA in order to measure students’ performance, cognitive states and satisfaction related to both assessment methods. The authors concluded that smaller versatile GBAs may have a greater impact on the student’s cognitive capabilities, and could enhance student performances during, for example, a final exam or short formative assessments. Moreover, Tsai et al. \cite{Tsai2015259} proposed an online learning system using different gaming modes of classic tic-tac-toe games to explore how different gaming modes and feedback types in this game-based formative assessment affect knowledge acquisition effectiveness and perceptions of participation.
    
    \item \texttt{Final user}: if the paper's results are to be used by final users or are validating the GBA, we consider that the main stakeholder will be the final user in that context (e.g., teachers and students). In their work, Ciman et al. \cite{Ciman20181703} designed a game to support children with cerebral visual impairment, developing a mobile version of the game to be used by children easily at home on any platform. Delcker \& Ifenthaler \cite{Delcker2020195} also developed a mobile app that makes an automated analysis of the data and provides information about children's language skills. Other papers focused on teachers, such as \cite{Martinez202097}, where the authors used a GBA to develop a set of visualizations to support teachers in classrooms.
\end{enumerate}

\begin{figure}[ht]
\includegraphics[width=0.9\textwidth]{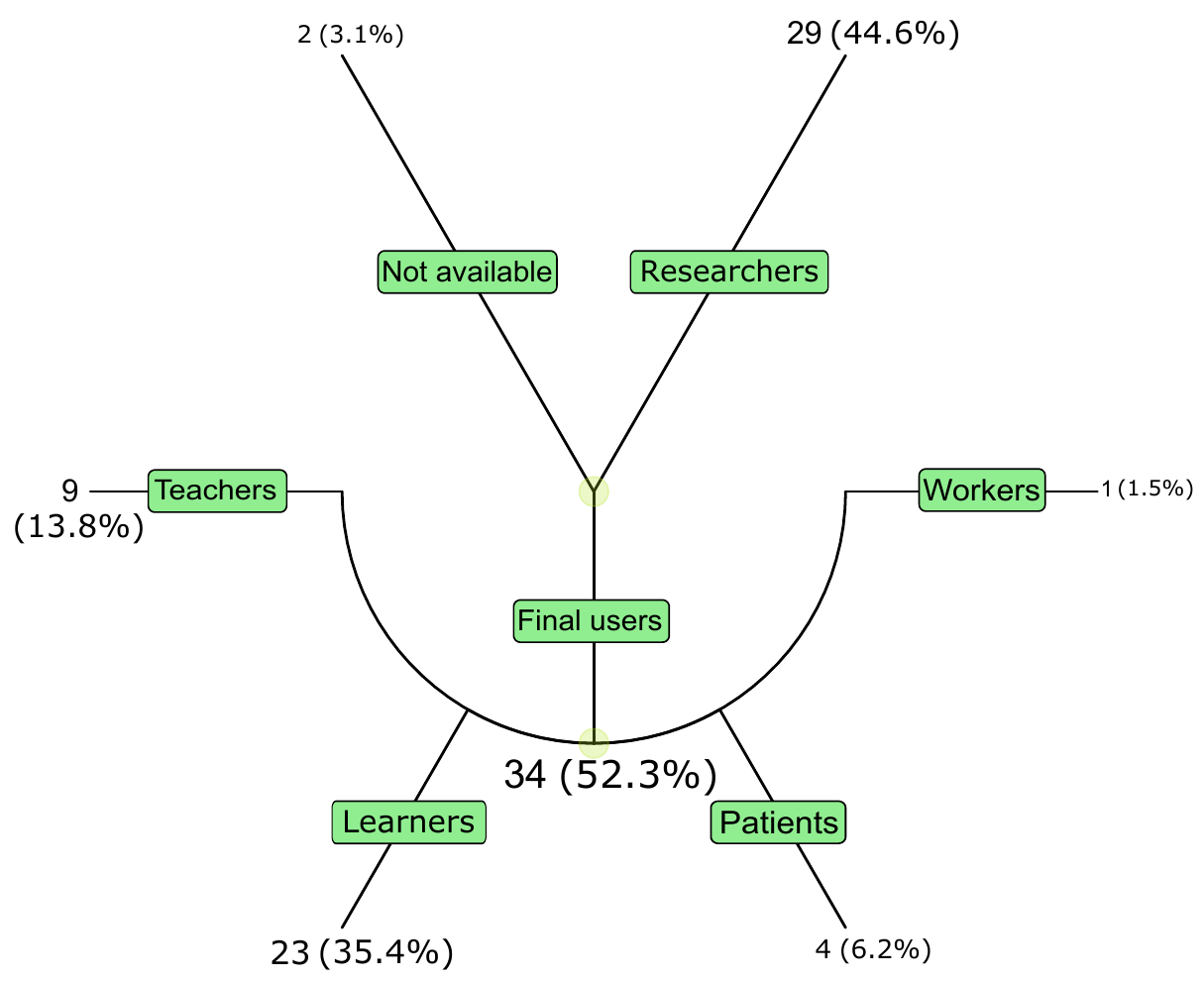} 
\centering
\caption{Paper distribution tree based on the main stakeholder.}
\label{fig:rq6}
\end{figure}

Figure \ref{fig:rq6} shows the number of studies that focused their work on each of the stakeholders. We see that 34 studies (52.3\%) were directed to \texttt{final users}, mainly students. Moreover, 29 studies (44.6\%) focused on \texttt{researchers} as the main result recipients. Two further studies (e.g., \cite{Paiva2018}) did not provide results and we categorized them as \texttt{NA}. Focusing on studies directed to the final user, we see that the majority of papers are directed to \texttt{learners} (35.4\%) and \texttt{teachers} (13.8\%).

\subsection{What limitations and challenges do the authors address? (RQ8)}
\label{subsec:rq8}

Limitations show potential weak points of the study that researchers usually highlight regarding their work such as constraints in research design or methodology. We can group the limitations that the authors faced in the six following categories: \texttt{game design}, \texttt{data sample}, \texttt{methodological}, \texttt{technical}, \texttt{integration} and \texttt{validation}.

\begin{enumerate}

     \item \texttt{Game design}: an appropriate design of a game is crucial for learners’ assessment since the GBA design must be adapted based on the constructs that will be evaluated. It requires a great effort to design a good GBA, aligning the evidence collected with the final purpose of the assessment. Designers might have different goals when developing a GBA \cite{Mislevy201523}: ``for game design, engagement; for instructional design, developing key concepts and capabilities in the target domain; for assessment design, evoking evidence of those capabilities for the intended use case(s).'' Moreover, many game design decisions play a role in what kind of game performance is achieved and its meaning \cite{Harteveld20152235}. Future designers should consider these concerns to achieve better designs, thus, creating more engagement and facilitating the development of the key concepts and capabilities intended for learning.
    
    \item \texttt{Data sample}: data were crucial for our review because GBA is based on the evidence, stored as data, generated by the students' interaction with the games. We examined each paper and found several limitations related to data. Jackson et al. \cite{Jackson2016796} reported that they had a small sample size and that larger sample sizes would be necessary to detect smaller effects. We see a similar example in \cite{Lehman2017917}, where the authors had a sample collection of 67 students, but only four of those 67 student samples were used in their empirical study. The work in \cite{Hao2018} described the difficulty of designing a good data model, as there are usually conflicts between programmers and assessment designers, usually complicated by constraints related to budgets and schedules.
    
    \item \texttt{Methodological}: this category includes challenges and limitations related to the methods, algorithms, or techniques used. For example, Yu et al. \cite{Yu2016727} wanted to collect additional data to explore learners' behavioral patterns during gameplay. We see another example in \cite{Shih20191255} since the authors reported that the assessment developed in this study only includes a part of number sense (this term refers to a group of key math abilities), and, in order to complete the number-sense battery, the assessment tools for the other components of number sense are needed to be developed.

    \item \texttt{Technical}: it is defined as a challenge involving how a machine or system works. This could include storage limitations, computing power, or even limitations related to sensors used in the study. In \cite{Ibryamova2020106}, the authors pointed out the necessity of a database (to store information about students' achievements), since they could not store that information, as well as the necessity of an administrator module to facilitate developing and modifying game elements.
    
    \item \texttt{Integration}: incorporating game activities as part of the curriculum in schools remains limited due to certain factors such as the schools' budget or the rigidity of subjects' classic curriculum. Halverson \& Owen \cite{Halverson2014111} claimed that if GBAs can show that games can serve as assessments that generate reliable evidence, we could then legitimize the potential of games and then break the social conventions that limit the potential of learning and assessment technologies in schools.
    
    \item \texttt{Validation}: one of the most significant parts of the research is the validation of the results. Validation is intended to ensure that the proposed methods and the accomplished results proved satisfactory by conducting empirical experiments. Sanchez \& Langer \cite{Sanchez202055} suggested that the games used in their study were entertainment games, and further research could be oriented to validate their results with games designed for assessment purposes.
\end{enumerate}

\begin{figure}[ht]
\includegraphics[width=0.9\textwidth]{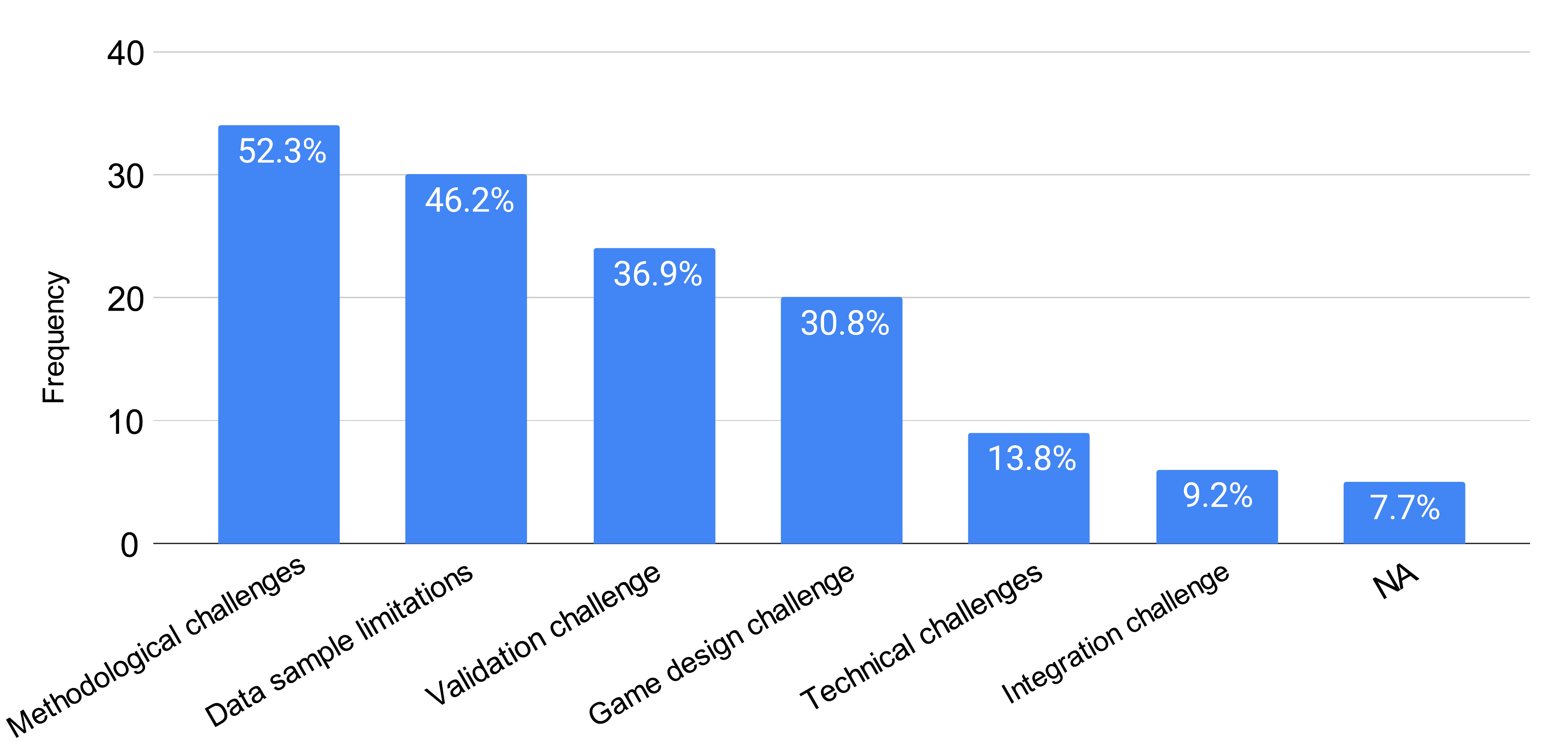} 
\centering
\caption{Number of papers based on their limitations and challenges.}
\label{fig:rq8}
\end{figure}

Some other studies did not report any challenges or limitations (e.g., \cite{Ponticorvo2017126}). Figure \ref{fig:rq8} shows that \texttt{methodological} challenges are the most common ones (34 studies, 52.3\%), followed by \texttt{data sample} limitations (30 studies, 46.2\%). On the other hand, the least frequent limitations are related to \texttt{integration} (6 studies, 9.2\%) and \texttt{technical challenges} (9 studies, 13.8\%).

\section{Discussion}
\label{sec:discussion}

In this section, we first present a summary and discussion of our main findings. We can also see a summary of these main findings in Table \ref{tab:summTable}. Next, we present a discussion of past and future challenges regarding games for assessment. Finally, we address existing limitations in this study and the implications of our research.

\begin{table}[t]
\centering
\begin{tabular}{| p{.3\textwidth} | p{.41\textwidth} | p{.08\textwidth}| p{.08\textwidth}|}
\hline

\textbf{Research question} & \textbf{Categories} & \textbf{Count} & \% \\ \hline

{\textbf{Context (RQ1)}} & Formal education & 41 & 63.1\% \\ \cline{2-4}

& Medical & 8 & 12.3\% \\ \cline{2-4}
 
& Workplace & 5 & 7.7\% \\ \cline{2-4}

& Not Available & 11 & 16.9\% \\ \hline

{\textbf{Main Purpose}} & GBA evaluation & 38 & 58.5\%  \\ \cline{2-4}

\textbf{(RQ2)} & In-game behaviors & 8 & 12.3\% \\ \cline{2-4}

& Assessment & 34 & 52.3\%   \\ \cline{2-4}

& Interventions & 7 & 10.8\% \\ \cline{2-4}

& Framework proposal & 12 & 18.5\%  \\ \cline{2-4}

& Game design proposal & 3 & 4.6\% \\ \hline

\textbf{Domain (RQ3)} & STEM & 19 & 29.2\% \\ \cline{2-4}

& Cognitive and soft skills & 28 & 43.1\%  \\ \cline{2-4}

& Humanities and social sciences & 17 & 26.2\%  \\ \cline{2-4}

& Physiological capacities & 8 & 12.3\% \\ \cline{2-4}

& Not Available & 2 & 3.1\% \\ \hline

\textbf{Availability (RQ4)} & Not specified & 51 & 78.5\%  \\ \cline{2-4} 

& Currently available & 9 & 13.8\%  \\ \cline{2-4}

& Not available & 5 & 7.7\% \\ \hline

\textbf{Sample Size (RQ5)} & Fewer than 50 participants & 19 & 29.2\%  \\ \cline{2-4}

& Between 50 and 250 participants & 19 & 29.2\%  \\ \cline{2-4}

& Between 250 and 500 participants & 7 & 10.8\%  \\ \cline{2-4}

& More than 500 participants & 9 & 13.8\% \\ \cline{2-4}

& Not available & 11 & 16.9\% \\ \cline{2-4}

\hline

\textbf{Algorithms/} & Descriptive statistics & 54 & 83.1\%  \\ \cline{2-4}

\textbf{techniques (RQ6)} & Machine learning & 24 & 36.9\%  \\ \cline{2-4}

& Deep learning & 1 & 1.5\%  \\ \cline{2-4}

& Sequence mining & 1 & 1.5\%  \\ \cline{2-4}

& Knowledge inference & 7 & 10.8\% \\ \cline{2-4}

& Not available & 8 & 12.3\% \\ \hline

\textbf{Stakeholder (RQ7)} & Researchers & 29 & 44.6\%  \\ \cline{2-4}

& Final user & 34 & 52.3\% \\ \cline{2-4}

& Not available & 2 & 3.1\% \\ \hline

\textbf{Limitations (RQ8)} & Technical & 9 & 13.8\%  \\ \cline{2-4}

& Game design & 20 & 30.8\%  \\ \cline{2-4}

& Data sample & 30 & 46.2\%  \\ \cline{2-4}

& Methodological & 34 & 52.3\%  \\ \cline{2-4}

& Integration & 6 & 9.2\%  \\ \cline{2-4}

& Validation & 24 & 36.9\% \\ \cline{2-4}

& Not available & 5 & 7.7\% \\ \hline
\end{tabular}
\caption{Summary of the main findings.}
\label{tab:summTable}
\end{table}

% Discusión sobre resultados principales
\subsection{Current trends}

First of all, we analyzed the contexts where the studies were applied (RQ1), finding that most of them took place in \texttt{K-16 education}, especially in \texttt{high school and middle school}. This is an exciting finding because young kids and teenagers represent the major force whose 21st-century competencies development will be heavily impacted by technology \cite{qian2016game, robberts2022design}. Moreover, children and adolescents are an ideal target since the familiarity of these users with gaming environments and game mechanics facilitates their interactions with games \cite{alonso2019applications}.

Regarding RQ2, we found that the majority of GBA studies focused on students' \texttt{assessment} and the \texttt{validation} of the game or the tool used. This suggests that having established that games are helpful for other purposes beyond entertainment, there is an increasing interest in using games as a natural alternative to classic evaluation methods, validating and comparing them against those traditional alternatives. Moreover, the fact that researchers also focused on the validation of the GBA used is a promising finding. Specifically, Gris and Bengtson \cite{gris2021assessment} pointed out the lack of evidence about engagement and usability needs, especially with well-assessed reliability and validity. We also noticed that few studies had the main purpose of proposing or validating a \texttt{game design} for assessment. Although many studies proved that GBA could improve students' learning outcomes, we should not forget game design. The literature reveals that game design is essential, and several distinctive design elements, such as narrative context, rules, goals, rewards, multi-sensory cues, and interactivity, seem necessary to stimulate the desired outcomes \cite{dondlinger2007educational, gee2006video}.

We also extracted four predominant domains (RQ3) across studies. A large proportion of the analyzed studies aimed at practicing and assessing content related to \texttt{STEM} as well as \texttt{humanities and social sciences}. This is not surprising since many of the studies took place in schools and high schools, and the use of games in these contexts is an ideal opportunity to teach content related to the main subjects at those ages. Another large number of papers also focused on developing and measuring \texttt{cognitive and soft skills}. Using game design as a context to teach higher-order thinking skills has drawn attention from researchers since schools usually place heavy emphasis on covering and delivering content knowledge \cite{akcaoglu2014cognitive}. Moreover, this could be useful not only in educational contexts, as we have seen some studies that measure cognitive skills for \texttt{medical} purposes (e.g., rehabilitation) \cite{Vallejo2017, Loachamin-Valencia2017106}. However, the researchers in \cite{hussein2021digital} pointed out the lack of research on 21st-century skills such as creativity and critical thinking.

We discovered that many of the studies had small data samples (RQ5). Furthermore, a significant part of the studies did not specify the data sample size used in the experiment. This is also noticed by the researchers themselves, as nearly half of the studies reported data sample limitations. Moreover, apart from collecting the sample size, we also tried to collect information about the type of data collected. However, almost no study included information related to the type or format of the data used.

Across papers, researchers used many different algorithms and techniques (RQ6) to analyze the data. We classified them into five categories and found that the most common ones are \texttt{descriptive statistics} and \texttt{machine learning}. With that said, we note that the majority of papers used statistical analyses or basic machine learning algorithms, and few studies used more complex or advanced methods, which  might be more adequate to model students' knowledge properly. However, those techniques that are easier to implement are also the ones chosen more frequently by researchers. Therefore, more work is needed to develop specialized GBA methods, that are also affordable to implement. This could perhaps be done through open-source libraries and more reproducible research. Moreover, we consider that making results interpretable is an essential part of the assessment, and one way to reach this interpretability is by using \texttt{visualizations}. Visualizations are essential components of research presentation and communication because of their ability to represent large amounts of data \cite{ware2019information} and because it is easier for the brain to comprehend an image as opposed to words or numbers \cite{cukier2010special}. We think this is a promising way to integrate games in schools, and we realized that studies now tend to use visualizations to communicate their results (e.g., \cite{Song2020, Cutumisu20192977}).

Finally, we wish to report the scarce information regarding games and tools availability (RQ4). The majority of studies did not provide any information on how to access or use their tools. In addition, some studies made tools public but expired, being inaccessible at present. This underscores the low transference of this research to practice and, thus, we encourage authors to make their products and results publicly accessible since we consider that this is an essential part of this type of research.

\subsection{Open challenges}

From our results and previous related reviews, we find some open challenges in the area that authors usually report. A description of each of these challenges is found below:

In \cite{chin2009assessment}, the authors address the challenge of how to make appropriate assessments. They noted that pre-test and post-test measures are a good manner to make an assessment, and they also recommended unobtrusive ways to collect data, such as another person taking notes during game-play. In our review, we noted that, at present, most studies found an appropriate method to make good assessments using Evidence-centered Design (ECD) and stealth assessment. ECD framework views assessment as an evidentiary argument, that is, an argument from which we observe what students say, do, or make in a few particular circumstances \cite{mislevy2006implications}. Moreover, stealth assessment represents a unobtrusive, yet powerful process by which learner performance data are continuously gathered during playing and learning, and inferences are made about the level of relevant competencies, maintaining the learners' flow and engagement \cite{shute2011stealth}. Since ECD and stealth assessment are two common practices in current research, we could claim that the objective of making appropriate assessments using unobtrusive methods has been accomplished.

What data are going to be collected is as important as how to collect these data, and another present challenge is the design of games for specific assessment purposes. Akcaoglu \& Koehler \cite{akcaoglu2014cognitive} indicated that games that present a hidden questionnaire format do not engage learners, while well-designed games can engage learners in reflective thinking \cite{JOHNSON20101246}. Although we identified a few papers with the main objective of providing a good game design, many of them have developed an excellent game for other purposes. Some examples are \cite{Ruiperez-Valiente2020648, Vallejo2017}, and \cite{Abeele2015331}. Future research should focus on complex game designs rather than the typical simple quiz design, employing multiple game-design elements such as collaboration, role-playing, narrative, exploration and complexity \cite{qian2016game}.

An important open challenge at present is replication and transferring the research to practice. In addition to the findings in our literature review about the game or tool being unavailable in most cases, All et al. \cite{all2014measuring} mentioned replication issues with certain studies due to missing information in multiple areas of the study. It is crucial to provide a detailed description of the procedure followed to conduct the study. While the community is currently demanding more standardized open science practices, this problem is still present currently. Besides, Alonso-Fernandez et al. \cite{alonso2019applications} noted that most papers did not describe the format in which they collected the data, so we cannot know if they used a standard or relied on their data formats, which represent even more replication and reusability issues. In addition, having open-source games or tools would be especially helpful for researchers. Unfortunately, we did not find any available open-source games across the studies. This problem of missing information is a familiar issue in multiple research fields (nearly every field is affected), and it leads to other problems such as low reproducibility. In fact, the terms ``reproducibility crisis'' and ``replication crisis'' have gained significant popularity over the last decade \cite{fidler2018reproducibility}. To fix this issue, the community is demanding more pre-registered studies, open data, open analyses, and open access publications \cite{van2018open}, and this can be systematized by the guidelines of the publishers, governments and research communities \cite{buck2015solving}.

Regarding the methods and techniques, we identify the challenge of implementing learner modeling algorithms. As we mentioned above, researchers usually use simple techniques to conduct their studies. In addition to Alonso-Fernandez et al. \cite{alonso2019applications} noting that limitation, we confirmed it in our results. In our review, 52.3\% of the papers reported methodological challenges to be addressed in future research, most of them related to the use of more complex metrics and techniques to infer new information. It is important to benefit from more advanced techniques (e.g., knowledge inference techniques, deep learning techniques) that can allow us to infer more complex and valuable information from the data collected. However, an important limitation of many of those advanced techniques is their low interpretability. Even if visualizations are a promising way to improve the presentation of results and communication, they cannot improve the model's interpretability themselves. According to the researchers in \cite{poursabzi2021manipulating}, with machine learning models being increasingly used, there has been an interest in developing interpretable models. However, there have been relatively few experimental studies investigating whether these models achieve their intended effects. Thus, the development of new models to provide better interpretability in GBA environments and their validation is still an open challenge.

We found several studies that described data sample and validation challenges. Since most evaluations are conducted with small samples, typically corresponding to one classroom's size, these studies present low statistical power, having a reduced chance of detecting actual effects \cite{petri2017games}. Thus, studies must use larger data samples to improve the results' generalization and validity. However, collecting large samples of in-context data is also a cumbersome task. Finally, a few empirical studies discussed the challenge of implementing GBA in the classroom, but this is a significant problem. Many teachers are still unsure about how to integrate game activities with the regular curriculum, and it is crucial to provide guidelines that can facilitate teachers to deploy games in the classroom more easily and flexibly \cite{gomez2021applying}.

\subsection{Limitations and implications}
% Motivando de nuevo por qué es importante: Implicaciones de GBA en educación/workforce 

This review is mainly limited by the paper selection. First of all, we have only used the key term ``game-based assessment'' to perform our document search, based on the papers' keywords and titles. However, other communities could also be working on games for assessment purposes, but they might be using slightly different terms to describe their work. Therefore, those studies might not be included in our review. Nevertheless, we purposely opted for this term to analyze the core of GBA while also having a manageable selection of papers for this study. Furthermore, we focused our attention on Scopus and the Web of Science, the two primary academic databases. However, there could be other peer-reviewed academic papers indexed in different databases, as well as non-peer-reviewed publications including pre-prints, technical or white reports that could be missing in our review, and also non-academic work being conducted in industrial companies and by practitioners. Regarding the computational methods and algorithms used, we have identified a set of categories based on the qualitative review of each selected paper. However, there might be studies using less quantitative approaches that might be missing in this review due to the review methodology itself. Finally, we have based our RQ generation on a simplified process that involves the general steps required in GBA projects, but there might be other potential and valuable RQs about the GBA field missing in this review.

We found that most studies emphasized GBA implementation and comparisons between games and classic assessment methods. More studies are needed to systematically develop and improve game design, adopting design-based research methods, as mentioned in \cite{li2013game}. The potential of GBA is now emerging, coinciding with the rise of big data. Data mining and visualization techniques on player interaction logs can provide different stakeholders with valuable insights into how players interact with the game \cite{freire2016game}. The increasing interest in games as a learning tool also indicates their potential as actual assessment tools. In our review, we found that GBAs are not only being used in K-16 education but also in medical and professional areas, among others. As expected, the most frequent area where GBAs are being applied is \texttt{K-16 education} since children and adolescents are the leading groups whose development will be affected by technology. 

Despite this dominating use in education, we can see the great potential that GBAs have in many other contexts. Concerning the professional environment, companies have begun to include assessment games for the recruitment of staff and the selection process. This is a relatively new trend due to certain limitations, such as the cross-domain generalizability of behaviors between game and workforce contexts, which needs further research \cite{Short2019161}. In medical environments, the use of GBAs can also be helpful for multiple purposes. Some examples are the possibility to recreate a virtual environment with daily life activities, allowing a precise and complete cognitive evaluation, which can be useful to treat certain diseases such as Alzheimer's \cite{Vallejo2017} or using games to rehabilitate children with cerebral visual impairment using an eye-tracker \cite{Ciman20181703}. Due to the above, we firmly believe that the future of games for assessment is promising; however, further research is needed to overcome the existing problems, and increase the still limited application of games in real-life environments, in order to start building the classrooms of the future.

\section{Conclusions}
\label{sec:conclusion}

Technology is changing and improving every day, and this is also making a significant impact on educational areas. Moreover, playing games is one of the most popular activities in the world, and the technological revolution that we are experiencing allows the implementation of games as alternative assessment tools in educational environments. However, previous studies suggest that the use of games also presents some challenges, such as finding the time for both the presenter/instructor and student to learn the systems employed, the financial impact on both parties, and technical limitations \cite{eiland2019considerations, de2017future}. We can tackle all these challenges by facing current limitations and revealing the great potential games have for assessment. This study represents a novel analysis and the first literature review of the emerging research field of GBA. Its main purpose was to review empirical studies of digital GBAs published until 2020. Based on a detailed systematic review of the 65 selected papers, we concluded that games are mainly used in K-16 education for assessment and validation purposes. The domain of the games used is usually related to STEM and cognitive skills, but other domains emerged from our analysis, such as social sciences and physiological capacities. Moreover, we note that, although few GBA studies had the purpose of proposing an adequate game design for assessment, most studies used games designed specifically for assessment purposes, employing complex game-design elements such as collaboration, narrative, or role-playing. In addition, we found that most of the studies used small data samples and simple techniques to process these data and assess students. Finally, we found that most of the studies do not provide public access to their tools, or they overlook links and let them expire over time, which makes it impossible to reproduce the results or even try their game.

% Future work
Future work should address the current challenges emerging from our review, as those are the main barriers to actual systematic adoption of games for assessment. For example, the next generation of GBA studies should ensure that enough data is collected to have meaningful and reliable results since one of the main limitations of the current research was the size of the data sample collected. Moreover, they should also address the game design that will be used for assessment, as many studies use games designed for other purposes (e.g., entertainment) and overlook the vital link between the design of a game and collecting the necessary evidence for the assessment. In that sense, it would be good to work on conceptual GBA pieces or frameworks that can provide a set of guidelines for the design. Moreover, classic performance indicators such as completion times or scores could still be included in future studies, but GBA also needs to apply more specific and complex algorithms (e.g. knowledge inference or deep learning techniques) specifically designed for learner modeling and assessment purposes. The use of more complex techniques, along with larger data samples, could substantially improve the reliability and generalization of the results. We also believe that future studies should continue exploring the use of visualizations and dashboards to integrate games in schools, adopting a more intuitive approach rather than providing teachers with raw numerical outputs or metrics, which are usually harder to understand. Teachers should also have a more important role in future work to address digital and assessment literacy issues, as well as the potential interpretability and actionability of GBAs. Finally, there are no theoretical frameworks within the GBA area (a related one regarding serious games could be \cite{loh2015serious}). Considering this lack of theoretical papers focused on describing GBA foundations, we believe that future work should address publications with additional content on the theoretical side.

Therefore, further research is needed to overcome current limitations and to continue exploring the possibilities of games as assessment tools in other contexts and environments. Finally, we encourage authors to document their research in a reproducible and verifiable way, using beneficial open science practices by pre-registering their study, sharing data and code for replication purposes, and if possible open sourcing the GBA tools with clear descriptions so that they can be used by interested stakeholders and researchers.

%Bibliography
\bibliographystyle{unsrt}  
\bibliography{references}

\end{document}